      [1999/12/01 v1.4c Il Nuovo Cimento]
\documentclass{cimento}
\usepackage{graphicx}

\title{On the spectra and distance scale of short Gamma Ray--Bursts}
\author{G.~Ghirlanda\from{ins:x}}
\instlist{\inst{ins:x} INAF - Osservatorio Astronomico di Brera}
\PACSes{\PACSit{98.70.Rz - 95.85.Pw - 95.75.Fg}{}}
\begin{document}

\maketitle

\begin{abstract}
I discuss the spectral properties of short GRBs detected by BATSE and
compare them with long events. The detailed modeling of short GRB
spectra shows that their being (on average) harder than long events
(as it has been found by comparing their hardness ratios) is indeed
due to a harder low energy spectral component (i.e. the powerlaw of
the Band or cutoff-powerlaw model) in short GRBs which, instead, have
a peak energy similar to long events. Another open issue is the
distance scale to short GRBs. While the few redshift measurements
suggests that they are at $z\le 1$, statistical studies (with local
galaxies or X--ray selected clusters) suggest that they should be even
more local.
\end{abstract}

\section{Short GRB spectra}
One of the still open issues of the population of short GRBs
(i.e. lasting less than 2 sec according to the BATSE duration
distribution) is the nature of their progenitors. The recent
discoveries of their afterglow emission (e.g. \cite{ref:ghe}) and the
first measurements of their redshifts enriched the limited
observational picture based on the study of their prompt emission
alone. However, the understanding of their emission mechanism and
progenitors still requires to study their prompt emission up to few
MeV energies.

In fact, from the light curves of short and long GRBs detected by
BATSE (in its 9 years of activity), it was discovered
(e.g. \cite{ref:kou,ref:tav,ref:cli}) that, on average, short GRBs are
spectrally harder than long events (i.e. the {\it Hardness--Duration
paradigm}). This result was based on the hardness ratio (i.e. the
ratio of the fluence detected in two broad contiguous energy bands)
which, however, is a rough indicator of the real spectral properties
of GRBs. Indeed, it has been shown (e.g. \cite{ref:ban}) that GRB
spectra are described by different spectral slopes in different energy
ranges and that they often present a strong spectral evolution
(e.g. \cite{ref:pre,ref:ghi02}). 

\begin{figure}
\resizebox{13.5cm}{10cm}{\includegraphics{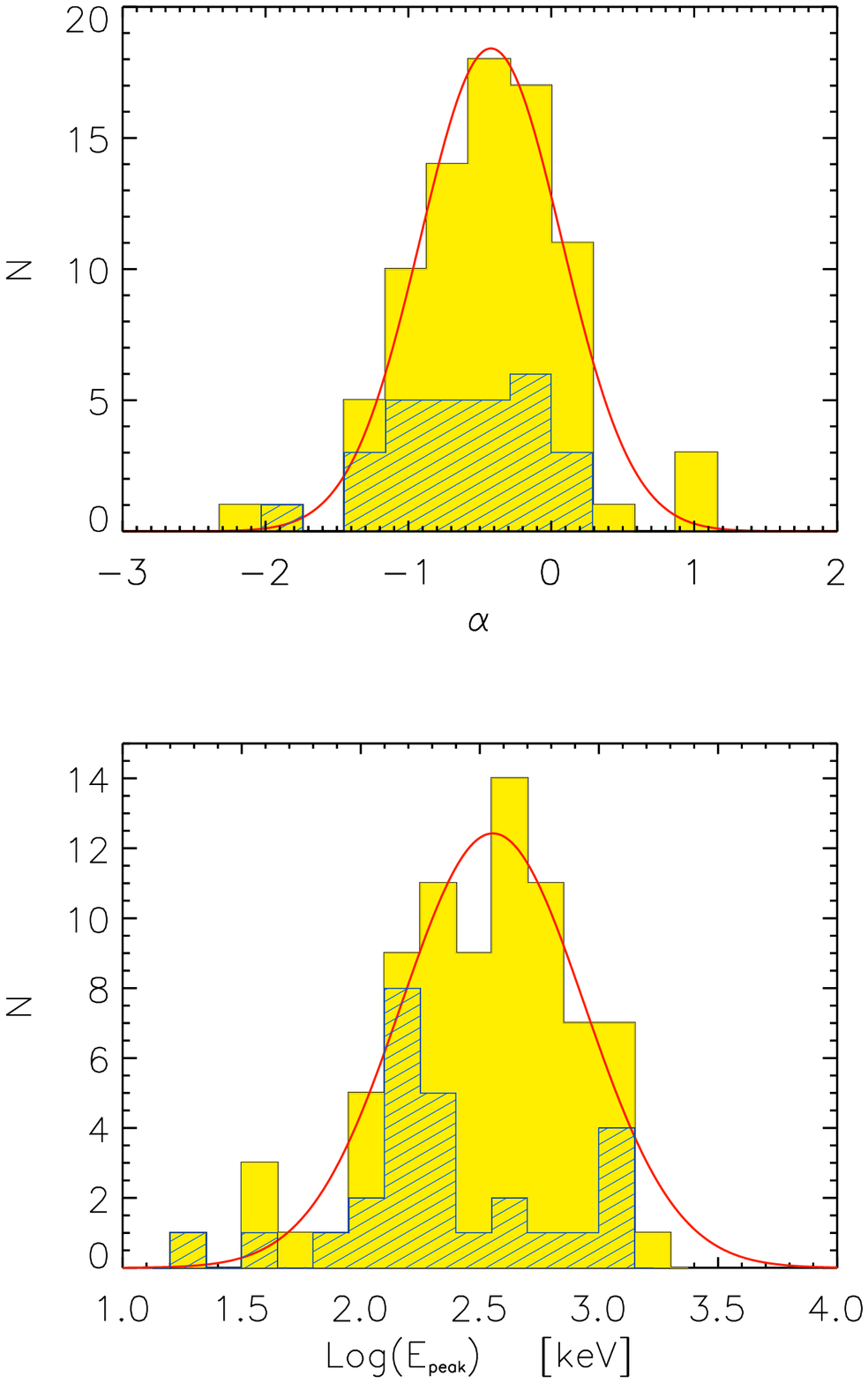}
\includegraphics{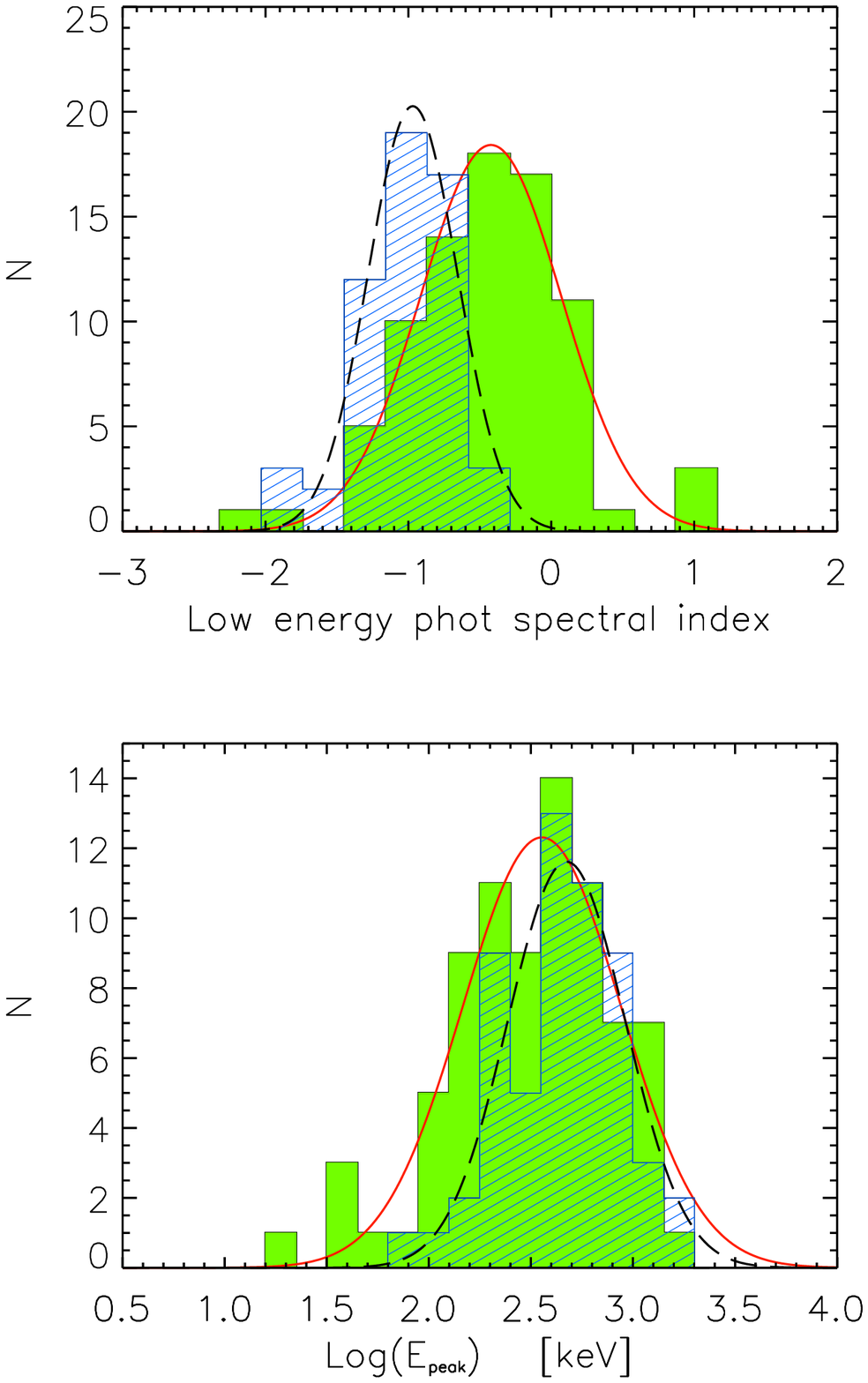} 
}
\caption{{\bf Left}: Spectral parameters
 ($\alpha$ and $E_{\rm p}$ of the cutoff--powerlaw model fit) of short
GRBs (\cite{ref:ghi}). The solid filled distributions refer to the
present sample of 81 GRBs, the hatched distribution is for the smaller
sample of 28 bright short GRBs (\cite{ref:ghi04}). {\bf Right}:
Spectral parameter distributions of short GRBs (the present sample -
solid filled histogram) compared to the sample of bright long GRBs
(from \cite{ref:ghi02}) fitted with the same model
(\cite{ref:ghi}). The Gaussian fits to these distributions are also
shown.}\label{spec}
\end{figure}

A detailed spectral analysis, limited to 28 bright short BATSE GRBs
(\cite{ref:ghi04}) with peak flux larger than 10
phot cm$^{-2}$ sec$^{-1}$, showed that their spectra are harder than
those of long bright GRBs (from \cite{ref:ghi02,ref:pre}) {\it due to
a harder low energy spectral component} rather than to a harder peak
energy. Moreover, this analysis indicated  that the spectra of
short GRBs are similar to the spectra of the first 2 sec of emission
of long events. 

We recently extended the spectral analysis to short GRBs with a lower
peak flux cut, i.e. 3 phot cm$^{-2}$ sec$^{-1}$ (in the 50--300 keV
energy range). Only in 81 (out of 157 selected) bursts we could
properly constrain the spectral parameters and in most cases the
spectrum was better fitted with an exponential cutoff powerlaw
model. However, due to the low signal--to--noise of the high energy
spectrum (above $\sim$300 keV), it cannot be concluded that the
preference for a cutoff--powerlaw model fit is a characteristic of
short GRB spectra ().

These results (Fig.~\ref{spec} - left panels) indicate that the
spectral parameters distribution (i.e. low energy photon spectral
index $\alpha$ and peak energy $E_{\rm p}$) of the bright short bursts
and of the larger sample (extended to lower peak fluxes) are similar.
When compared with the population of long GRBs\footnote{Note that the
comparison sample is with bright long GRBs. The analysis of a larger
sample of long GRBs with a similar peak flux cut is in progress
(Nava et al. in prep.).} (Fig.~\ref{spec} - right panels) it is confirmed
that the Hardness-Duration paradigm is due to a harder low energy
spectral component in short GRBs with respect to long events, while
the peak energy distributions of short and long GRBs are similar.

\section{Redshifts}

Another still open issue related to short GRBs is their distance
scale. On the one side the few redshifts measured so far suggest that
they are relatively local (i.e at $z \le 1$, see \cite{ref:ber}) while
on the other side there are statistical results, based on the BATSE short
GRB population, that indicate that short GRBs should be even more
local (\cite{ref:tan}). The problem of the distance scale reflects on
the energetic of short bursts and also on their progenitor nature. As
a consequence of the double compact evolved star merger model for
short bursts, it was expected to find them preferentially associated
to evolved ellipticals. However, there are growing theoretical and
observational evidences showing that short bursts can also be found in
late type star forming galaxies (e.g. \cite{ref:cov}). Few recent
cases have also been found in cluster members
(e.g. \cite{ref:blo,ref:gla}).

\begin{figure}
\resizebox{13cm}{6cm}{\includegraphics{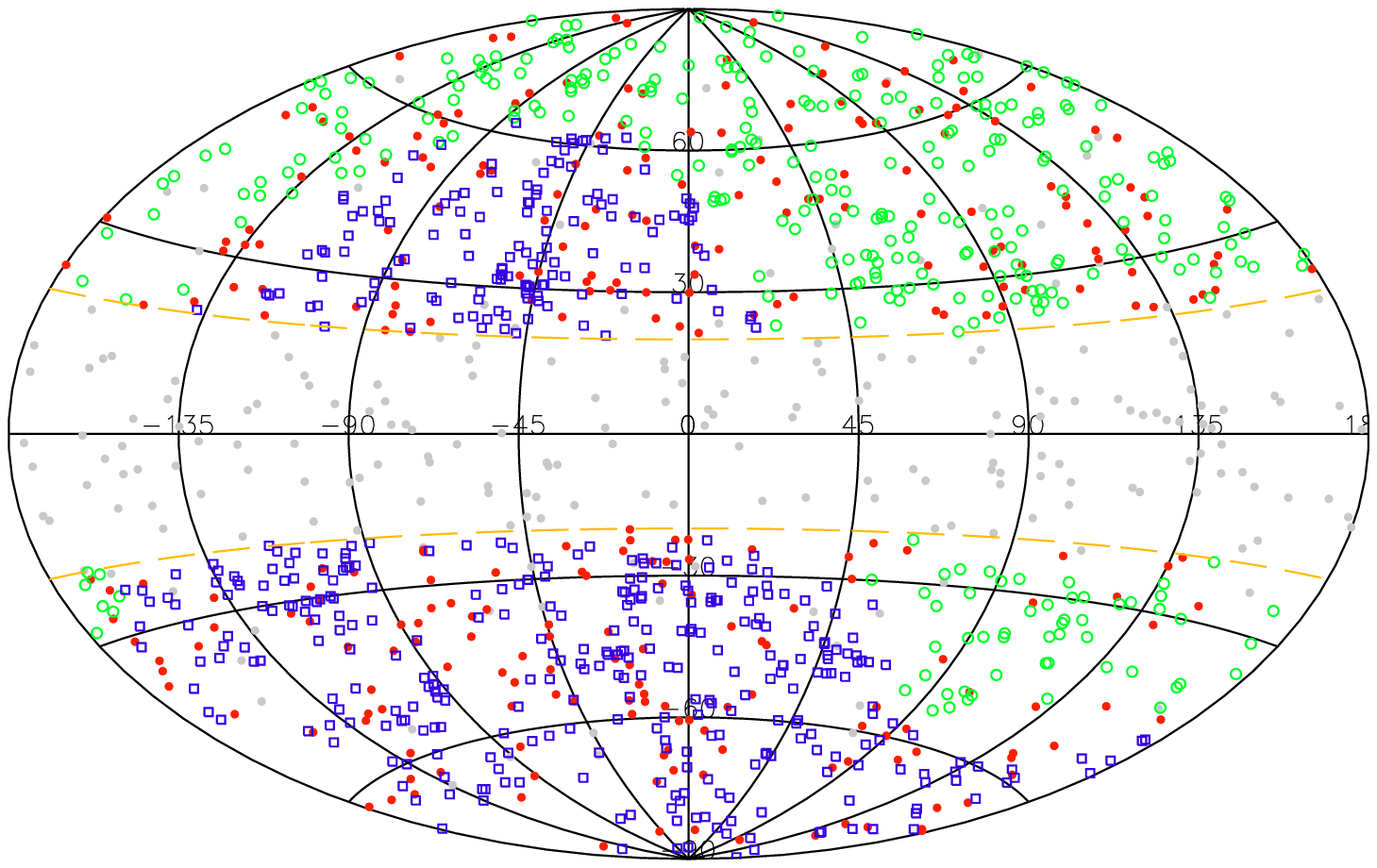}
\includegraphics{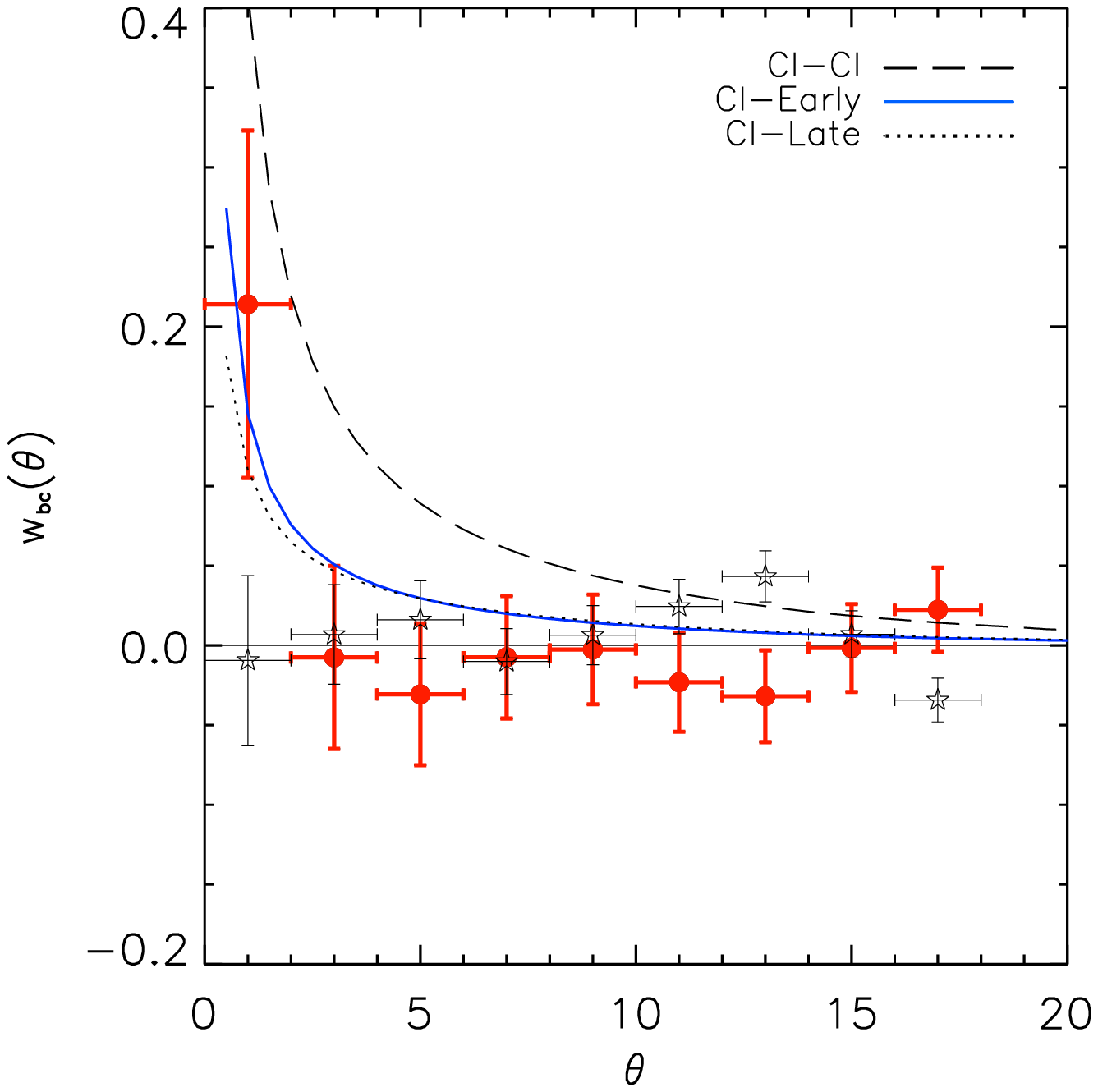} 
}   
\caption{{\bf Left}: Sky distribution in Galactic coordinates of the 
sample of 497 BATSE short GRBs (filled circles) and of 763 Clusters
(449 REFLEX clusters - blue open squares - and 314 NORAS clusters -
green open squares). The red filled circles represent the 283 short
duration GRBs with position accuracy $<10^{\circ}$ and
$|b|>20^{\circ}$ which has been used to compute the cross correlation
analysis with the sample of Clusters (see \cite{ref:ghi06a}). {\bf
Right}: Angular cross-correlation (red filled circles) between short
GRBs and X---ray selected clusters of galaxies. It is also shown
(open stars) the signal obtained with BATSE long GRBs. The
long--dashed line represents the autocorrelation function of clusters,
while the solid and dotted lines represent the cross correlation of
clusters with early type and late type galaxies,
respectively. }\label{sky}
\end{figure}

We studied (\cite{ref:ghi06a}) the possible correlation between BATSE
GRBs (497 short and 1540 long events) and X-ray selected clusters of
galaxies (763, 314 from the NORAS and 484 from the REFLEX samples -
\cite{ref:boe,ref:boe1}). The results indicate that short GRBs 
are correlated with clusters while we do not find any correlation with
the population of long GRBs. Moreover, by comparing the burst--cluster
correlation function with the cluster--cluster auto--correlation
function we can exclude that short bursts trace the cluster
distribution. Instead, through the comparison with the cluster--galaxy
correlation functions we conclude that short GRBs are associated with
``normal'' galaxies.

By selecting local cluster sub--samples (or flux cuts in the flux
limited cluster sample) we find a higher correlation signal with local
($z<0.06$) or sub-luminous clusters. This result is in agreement and
further supported by the finding of a short GRB--cluster cross
correlation function similar to the galaxies--cluster
cross--correlation function and the fact that the short GRBs
auto--correlation function is  similar to that of local ($z<0.1$)
galaxies (\cite{ref:mag}).

These results are still a challenge for the (still) few spectroscopic
redshift measurements of short bursts. However, if correct they predict
a typical energy of short GRBs of $\sim 10^{48}$ erg, i.e. a factor
$10^{4}$ smaller than long GRBs.

\acknowledgments
I am grateful to A. Celotti, G. Ghisellini, L. Guzzo, M. Magliocchetti
for valuable collaboration.


\begin{thebibliography}{0}
\bibitem{ref:ghe} \BY{Gerhels~N. et al.}
  \IN{Nature}{437}{2005}{851}.
\bibitem{ref:kou} \BY{Kouveliotou~C. et al.}
  \IN{ApJ}{413}{1993}{L101}.
\bibitem{ref:tav} \BY{Tavani~M. et al.}
  \IN{ApJ}{497}{1998}{L21}.
\bibitem{ref:cli} \BY{Cline~B. D., Matthey C. \atque Otwinowski S.}
  \IN{ApJ}{527}{1999}{827}.
\bibitem{ref:ban} \BY{Band~D. L. et al.}
  \IN{ApJ}{413}{1993}{218}.
\bibitem{ref:pre} \BY{Preece~R. D. et al.}
  \IN{ApJS}{126}{2000}{19}.
\bibitem{ref:ghi} \BY{Ghirlanda et al.}
  \IN{A\&A}{}{2005}{Submitted}.
\bibitem{ref:ghi04} \BY{Ghirlanda~G., Ghisellini G. \atque Celotti A.}
  \IN{A\&A}{422}{2004}{L55}.
\bibitem{ref:ghi02} \BY{Ghirlanda~G., Celotti A. \atque Ghisellini G.}
  \IN{A\&A}{393}{2002}{409}.
\bibitem{ref:ber} \BY{Berger~E.}
  \IN{``Gamma Ray Bursts in the Swift Era'', eds. S. Holt, N. Gehrels
  and J. Nousek}{2006}{astro-ph/0602004}.
\bibitem{ref:tan} \BY{Tanvir~N. et al.}
  \IN{Nature}{438}{2005}{991}.
\bibitem{ref:cov} \BY{Covino~S. et al.}
  \IN{GCN}{}{2005}{3665}.
\bibitem{ref:blo} \BY{Bloom~J. et al.}
  \IN{ApJ}{638}{2005}{354}.
\bibitem{ref:gla} \BY{Gladders~G. et al.}
  \IN{GCN}{}{2005}{3798}.
\bibitem{ref:ghi06a} \BY{Ghirlanda~G. et al.}
  \IN{MNRAS}{368}{2006}{L20}.
\bibitem{ref:boe} \BY{Boeringher~H. et al.}
  \IN{ApJSS}{129}{2004}{435}.
\bibitem{ref:boe1} \BY{Boeringher~H. et al.}
  \IN{A\&A}{425}{2006}{367}.
\bibitem{ref:mag} \BY{Magliocchetti~M., Ghirlanda G. \atque Celotti A.}
  \IN{MNRAS}{343}{2003}{255}.
\end{thebibliography}
\end{document}